\documentclass[pra,preprint,superscriptaddress,amsmath,amssymb,floatfix,showpacs]{revtex4}

\usepackage{graphicx}
\usepackage{epsfig}
\usepackage{wrapfig}
\usepackage{braket}

\usepackage[tableposition=top]{caption}

\begin{document} 
\title{The electronic structure of palladium in the presence of many-body effects}
\author{A. \"Ostlin}
\affiliation{Theoretical Physics III, Center for Electronic
Correlations and Magnetism, Institute of Physics, University of
Augsburg, D-86135 Augsburg, Germany}
\affiliation{Department of Materials Science and Engineering, Applied Materials Physics, KTH Royal Institute of Technology, SE-10044 Stockholm, Sweden}
\author{W. H. Appelt}
\affiliation{Augsburg Center for Innovative Technologies, University of Augsburg,
D-86135 Augsburg, Germany}
\affiliation{Theoretical Physics III, Center for Electronic
Correlations and Magnetism, Institute of Physics, University of
Augsburg, D-86135 Augsburg, Germany}
\author{I. Di Marco}
\affiliation{Department of Physics and Astronomy, Division of Materials Theory,
Uppsala University, Box 516, SE-75120 Uppsala, Sweden}
\author{W. Sun}
\affiliation{Department of Physics and Astronomy, Division of Materials Theory,
Uppsala University, Box 516, SE-75120 Uppsala, Sweden}
\author{M.~Radonjic}
\affiliation{Theoretical Physics III, Center for Electronic
Correlations and Magnetism, Institute of Physics, University of
Augsburg, D-86135 Augsburg, Germany}
\affiliation{Scientific Computing Laboratory, Institute of Physics Belgrade, University of Belgrade, Pregrevica 118, 11080 Belgrade, Serbia}
\author{M.~Sekania}
\affiliation{Theoretical Physics III, Center for Electronic
Correlations and Magnetism, Institute of Physics, University of
Augsburg, D-86135 Augsburg, Germany}
\affiliation{Andronikashvili Institute of Physics, Tamarashvili 6, 0177 Tbilisi, Georgia}
\author{L. Vitos}
\affiliation{Department of Materials Science and Engineering, Applied Materials Physics, KTH Royal Institute of Technology, SE-10044 Stockholm, Sweden}
\affiliation{Department of Physics and Astronomy, Division of Materials Theory,
Uppsala University, Box 516, SE-75120 Uppsala, Sweden}
\affiliation{Research Institute for Solid State Physics and Optics, Wigner Research Center for Physics, P.O. Box 49, H-1525 Budapest, Hungary}
\author{O. Tjernberg}
\affiliation{KTH Royal Institute of Technology, Materials Physics, SE-16440 Kista, Sweden}
\author{L. Chioncel}
\affiliation{Augsburg Center for Innovative Technologies, University of Augsburg,
D-86135 Augsburg, Germany}
\affiliation{Theoretical Physics III, Center for Electronic
Correlations and Magnetism, Institute of Physics, University of
Augsburg, D-86135 Augsburg, Germany}

\begin{abstract} 
Including on-site electronic interactions described by the multi-orbital 
Hubbard model we study the correlation effects in the electronic structure 
of bulk palladium. We use a combined density functional and 
dynamical mean field theory, LDA+DMFT, based on the fluctuation exchange approximation. The agreement between the experimentally determined and
the theoretical lattice constant and bulk modulus is improved when correlation effects are included. 
It is found that correlations modify the Fermi surface around the neck at the $L$-point while the Fermi surface 
tube structures show little correlation effects. At the 
same time we discuss the possibility of satellite formation in the high energy 
binding region. Spectral functions obtained within the LDA+DMFT and $GW$ methods are compared to
discuss non-local correlation effects. For relatively weak interaction strength of the local
Coulomb and exchange parameters spectra from LDA+DMFT shows no major difference in
comparison to $GW$.
\end{abstract} 

\maketitle 
\section{Introduction}\label{intro}

Transition metals have their density of states characterized by a partially filled 
narrow $d$-band, superimposed on a broad free electron-like $sp$-band. The shape of the 
$d$-band especially in the $3d$ series is a consequence of the construction of the 
$d$-orbitals, as they overlap only to a limited extent with orbitals
on neighboring atoms and consequently the hopping integrals between $d$-orbitals
is small, as is the bandwidth. This points towards the importance of short range
strong Coulomb repulsion for the $3d$ elements. An additional ingredient in the 
$3d$ series is the appearance of magnetism. In a partially filled shell of a free atom
the exchange interaction between electrons favors the parallel alignment of electron
spins (Hund's rule). In solids electrons enter in extended states/orbitals 
so there is a competition between the kinetic energy of the electron which 
favors no spin alignment and the exchange interaction which favors spin alignment. 
If the band is narrow the energy gain from the exchange interaction may win and 
the spin alignment is favored. In that sense, the occurrence of magnetism in the 
$3d$ series is a consequence of the narrowness of the $3d$ band. A quantitative
theory to explain the electronic structure and hence the physical properties of $3d$-elements 
has been consistently developed during the last decades in the form of 
the combined density functional theory (DFT) and dynamical 
mean field theory (DMFT)~\cite{me.vo.89,ge.ko.96,ko.vo.04,ko.sa.06} which is generally 
referred to as the LDA+DMFT method~\cite{ko.sa.06,held.07} 
(LDA = local density approximation). In the LDA+DMFT scheme the LDA 
provides the \emph{ab initio} material dependent input (orbitals and hopping parameters),
while the DMFT solves the many-body problem for the local interactions. Therefore the LDA+DMFT approach
is able to compute, and even predict,  properties of correlated materials. 
Theoretical results obtained with LDA+DMFT can be compared with experimental data obtained,
for example, by photoemission spectroscopy (PES)~\cite{mina.11}. In particular, this technique measures 
spectral functions, i.e., the imaginary part of the one-particle Green function, and thus 
determines correlation induced shifts of the spectral weight. Indeed, most experimental 
investigations on the electronic structure of the $3d$ metal Ni rely on PES~\cite{ae.kr.96,sc.pr.96}. 
Braun et al.~\cite{br.mi.12} demonstrated  the importance of local correlations in 
Ni by exploiting the magnetic circular dichroism in bulk sensitive soft X-ray PES measurements. 
One of the dominant correlation effects observed in the PES data for Ni is the satellite 
peak situated at $6\,$eV below the Fermi level~\cite{sa.br.12,hi.kn.79}. This feature is 
not captured by LDA, but is well explained by LDA+DMFT~\cite{li.ka.01}. LDA+DMFT 
also reproduces the correct width of the occupied  $3d$ bands and the exchange 
splitting~\cite{li.ka.01,gr.ma.07,sa.br.12}. 

As LDA+DMFT is very successful for $3d$ elements, 
this motivates us to investigate the applicability of LDA+DMFT to $4d$ transition metal 
elements. Transition metals from the $4d$ series have larger bandwidths compared
to that of the $3d$ elements and correspondingly larger kinetic energies, which 
will favor an itinerant band-like picture over an atomic-like localized picture 
and somewhat weaker correlation effects. In our present study we focus on the $4d$ metal palladium.
Despite being in the same group as Ni in the periodic table, the physical properties
of Pd are very different, so a theoretical study including local and non-local correlation 
effects is particularly desirable.
Due to its interest in fundamental 
condensed matter theory, and its industrial use as a catalyst and for hydrogen storage, 
the electronic structure of Pd has been widely studied over the years. 
As a late $4d$ transition metal element, Pd is not far from the ferromagnetic 
instability: it has a high density of states at the Fermi level 
and a large Stoner enhancement in the magnetic susceptibility~\cite{ch.fo.68}. 
On expansion of the lattice constant Pd turns ferromagnetic, as shown by DFT
calculations~\cite{mo.ma.89}. Experimental studies involving PES have been used in the search for signatures of 
electronic correlations in Pd such as the existence of satellites in the spectral 
function~\cite{ch.kr.81,ni.la.81}. 
Liebsch~\cite{liebsch.79,liebsch.81} investigated the satellite formation mechanism 
in detail using many-body methods, pointing out the importance of taking electron-hole 
and hole-hole scattering into account by ladder-like summations in the $T$-matrix formulation. 
M{\aa}rtensson and Johansson predicted a satellite in Pd \cite{ma.jo.80}, placing the 
satellite at 8 eV binding energy, in good agreement with later experimental findings ($\sim 8.5$ eV) by 
Chandesris et al.~\cite{ch.kr.81}. The method employed in Ref. \cite{ma.jo.80} was 
semi-empirical, using thermodynamic input data. In this study we will discuss the 
satellite formation in Pd using \emph{ab inito} self-consistent state-of-art 
calculations as well.

Complementary information can be obtained from the analysis of the Fermi surface (FS).
Features of the Fermi surface can be experimentally probed by photoemission spectroscopy 
and de Haas-van Alphen (dHvA) measurements. The so called Kohn anomalies~\cite{kohn.59} may 
appear in the phonon dispersion relations of metals, arising from virtual scattering of conduction
electrons from state {\bf k} to {\bf k}$^\prime$ connected by nesting vectors $\textbf{q}$, making 
the determination of possible FS nesting of interest. The appearance of a Kohn anomaly in Pd is still
debated~\cite{stew.08,li.ya.11}, 

Palladium is perhaps the best studied high-susceptibility paramagnet
and played an important role in elucidating several aspects of the theory of
spin fluctuations. 
Among the elements, Pd is traditionally taken as the best candidate for
observing spin fluctuations because of its high electronic density of states
and large Stoner enhancement in the magnetic susceptibility. Specific heat
experiments~\cite{hs.re.81} showed a reduction in the electronic specific
heat coefficient of $7\%$ in a magnetic field of about 10 T suggesting
that strong spin fluctuations appear in Pd. The reduction of spin-fluctuation
contributions to the electronic specific heat at high magnetic fields is well
established theoretically by several works: Doniach and Engelsberg
 \cite{do.en.66}, Berk and Schrieffer~\cite{be.sc.66},
B\'eal-Monod and co-authors~\cite{be.ma.68,be.la.80}, and many others.
In their classical works, the Crabtree group
experimentally investigated the evidence of spin fluctuations in Pd by
measuring the cyclotron effective masses and the amplitude of the dHvA
effect as a function of the magnetic field~\cite{jo.cr.84a,jo.cr.84b}. 
These typical measurements provide
in principle information about spin-fluctuation contributions to the conduction
electron properties. While the former allows one to obtain information about
the density of states at the Fermi level, which determines the electronic
specific heat, the latter measures the difference in volume between the spin
up and spin-down Fermi surfaces, which determines the magnetization. The absence
of significant field dependence of the cyclotron effective mass and the spin
splitting factor~\cite{jo.cr.84a,jo.cr.84b} implies that the spin-fluctuation
contributions to the electronic specific heat and static spin susceptibility
$\chi=M/H$ are not appreciably affected by applied fields up to $\sim 13$ T. This
is consistent with the theoretical estimations made by Brinkmann and
Engelsberg~\cite{br.en.68} and Hertel et al.~\cite{he.ap.80} that magnetic fields
much larger than 13 T are required to suppress the spin fluctuations in Pd.
The magnetic properties and dynamical fluctuations in Pd were discussed
recently by Larson et al.~\cite{la.ma.04}. Highly accurate LDA
calculations were performed to estimate the parameters entering in
Moriya's spin-fluctuation theory \cite{mori.85}, in particular the Landau functional
for Pd was used to connect critical fluctuations beyond the local density
approximation with the band structure. It was pointed out~\cite{la.ma.04}
that the key parameter for the non-trivial properties of Pd is the 
mean-square amplitude of the spin fluctuations, which is a non-local quantity 
determined by the momentum dependent spin susceptibility in a large part
of the Brillouin zone, and therefore non-locality is expected to play a significant
role in the physical properties. It is one of the aims of this work
to identify local and non-local correlation effects on the
spectral function by comparing results obtained via LDA+DMFT and $GW$~\cite{hedi.65} methods.

The results presented here include the electronic structure, the Fermi 
surface and nesting vectors of Pd, and the satellite formation in the high binding energy region of 
the density of states. Most of our 
results have been obtained within the full potential linearized muffin-tin orbitals (FPLMTO) 
method implemented within 
the RSPt code~\cite{rspt_book}, which has previously proven to be able to
 accurately determine ground state quantities within LDA+DMFT for $3d$ transition
 metals~\cite{ma.mi.09,gr.ma.12}. Self-consistent
quasiparticle $GW$ calculations have also been performed~\cite{sc.ko.06.2,ko.sc.07}, 
which allows us to discuss the effect of non-local electronic correlations
in Pd.
The paper is organized as follows: Section~\ref{intro} is an introduction. In 
Section~\ref{compdet} we present computational methods and details of 
the calculations. Section~\ref{reseos} presents total energy data, from which 
we extract the optimal $U$ and $J$ values, to match the experimental and the
calculated equilibrium lattice parameters. We also present results concerning the onset of 
ferromagnetic long range order upon lattice expansion. In Section~\ref{resspec} 
the calculated spectral function of palladium is shown, and the relation to the 
photoemission satellite is discussed in detail. The effect of non-local correlations
is discussed in Section \ref{resloc}. 

\section{Computational methods and details}\label{compdet}

\subsection{The LDA+DMFT method}

Correlation effects in the valence Pd $4d$ orbitals were included
via an on-site electron-electron interaction in the form
$\frac{1}{2}\sum_{{i \{m, \sigma \} }} U_{mm'm''m'''}
c^{\dag}_{im\sigma}c^{\dag}_{im'\sigma'}c_{im'''\sigma'}c_{im''\sigma} $.
Here, $c_{im\sigma}/c^\dagger_{im\sigma}$ annihilates/creates an electron with 
spin $\sigma$ on the orbital $m$ at the lattice site $i$.
The Coulomb matrix elements $U_{mm'm''m'''}$ are expressed in the usual
way~\cite{im.fu.98} in terms of Slater integrals.
Since specific correlation effects are already
included in the local spin-density approximation (LDA), so-called
``double counted'' terms must be subtracted. 
To take this into account, we employed the interpolation double counting scheme~\cite{pe.ma.03}.
For the impurity solver a fluctuation exchange (FLEX)~\cite{bi.sc.89} 
type of approximation was used for the multiorbital case~\cite{li.ka.98,ka.li.99,po.ka.05}.
In contrast to the original formulation of                                          
FLEX ~\cite{bi.sc.89}, the spin-polarized $T$-matrix FLEX (SPTFLEX), used for
 the present calculations treats the particle-particle and the particle-hole channel
differently~\cite{li.ka.98,ka.li.99,po.ka.05}. While the particle-particle
processes are important for the renormalization of the effective
interaction, the particle-hole channel describes the interaction
of electrons with the spin-fluctuations, which represents one of the most
relevant correlation effects in Pd. In addition the advantage                                    
of such a computational scheme is that the Coulomb matrix elements can be
considered in a full spin and orbital rotationally invariant form, for realistic materials.

\subsection{The self-consistent quasiparticle $GW$ method}
In recent years, first-principle calculations involving the $GW$ approximation \cite{hedi.65}
are becoming more popular. In particular self-consistent $GW$
formulations are promising because they can more accurately calculate quantities like 
band gaps compared to ``one-shot'' $GW$ approaches \cite{ko.sc.07}.
In such methods the first step is to compute the band structure of the solid,
usually within DFT-LDA. Using the random phase approximation (RPA),
the density response function is then calculated and used to evaluate the dielectric
function and the screened Coulomb interaction $W$. The matrix elements of the 
self-energy are added as corrections to the LDA eigenvalues, and the
effective potential is self-consistently updated. In spite of 
the simplified formalism of calculation, compared to that of the full $GW$ scheme, a good agreement 
with experiment for several materials has been obtained \cite{ko.sc.07}.
In this study we employed the quasiparticle self-consistent $GW$ (QSGW) method~\cite{sc.ko.06.2,ko.sc.07}.
Our main object of interest is the self-energy corrected eigenvalue for band $n$ and Bloch vector \textbf{k},
\begin{equation}
E_{{\bf k}n} =  \epsilon_{{\bf k}n} + Z_{{\bf k}n} \Delta \Sigma_{{\bf k}n} 
\end{equation}
where the operator $\Delta \Sigma_{{\bf k}n} = \bra{\Psi_{{\bf k}n}} \Sigma({\bf r},{\bf r}^\prime,\epsilon_{{\bf k}n}) - V_{xc}({\bf r}) \ket{\Psi_{{\bf k}n}}$. The self-energy is given 
in terms of the Green's function and the screened Coulomb interaction $W$: 
$\Sigma({\bf r},{\bf r}^\prime, \omega)= \frac{i}{2\pi}\int d \omega^\prime G ({\bf r},{\bf r}^\prime, \omega-\omega^\prime)
W({\bf r},{\bf r}^\prime, \omega^\prime)e^{-\delta\omega^\prime}$. From the slope of the 
real part one can get the renormalization factor
\begin{equation}
Z_{{\bf k}n} = \left[ 1- \frac{\partial Re \Sigma_{{\bf k}n} (\omega)}{ \partial \omega} \right]^{-1}. 
\end{equation}
In a direct comparison with
the LDA+DMFT results, $GW$ calculations reveal if significant non-local correlation
effects occur in Pd.

\subsection{Technical details}
The LDA+DMFT calculations were done using the full-potential FPLMTO code RSPt~\cite{rspt_book} as a base for the underlying density functional theory calculations.
The RSPt calculations were based on the local-density approximation with the parametrization of Perdew
 and Wang \cite{pe.wa.92} for the exchange-correlation functional.
Three kinetic energy tails were used, with corresponding energies 0.3, -2.3 and -1.5 Ry. 
Palladium is a face-centered cubic metal, and the \textbf{k}-mesh we used had the size $16 \times 16 \times 16$ for the equations of state, $24 \times 24 \times 24$ for the other calculations, and Fermi-Dirac smearing with $T=400$ K (the same
temperature as was used for the imaginary frequency Matsubara mesh). The muffin-tin radius was set to 2.45 Bohr atomic
units (a.u.), and was kept constant throughout for all unit cell volumes. For the
charge density and potential angular decomposition inside the
muffin-tin spheres, a maximum angular momentum $l_{max} = 8$ was set. The calculations included spin-orbit coupling
and scalar-relativistic terms within the muffin-tin spheres, unless otherwise noted.
The SPTFLEX impurity solver was implemented in the
Matsubara domain, and we used 2048 imaginary frequencies and an electronic temperature 400 K. The analytic continuations of the self-energy from imaginary frequencies to the real energy axis in
the complex plane were performed by Pad\'e approximants \cite{vi.se.77}.

The QSGW scheme used in this study is implemented into the LMSuite package~\cite{sc.ko.06.2,ko.sc.07}, which is based on the full-potential linear muffin-tin orbitals code by M. Methfessel et al. \cite{me.sc.00}. The muffin-tin radius was chosen to be 2.63 a.u, and the integration of the Brillouin zone (BZ) was mapped with $24\times24\times24$ \textbf{k}-points. For the $GW$ calculation, we reduced the \textbf{k}-points to $6\times6\times6$~\cite{ko.sc.07}. A double-$\kappa$ basis set with $l_{max} = 4$ was used, including the semicore 4\textit{p} states with local orbitals. This basis set allows for an accurate description of the high-lying conduction band states. Spin-orbit coupling was included within the muffin-tin spheres.

We point out that both the RSPt and the QSGW methods are employing the full-potential linearized muffin-tin orbital basis set, but using different implementations. As can be seen in Section \ref{resloc}, this causes no major differences between the RSPt and the QSGW LDA level results.

\section{Results and discussion}\label{resdis}

\subsection{Equation of state}\label{reseos}

\begin{figure}[h!]
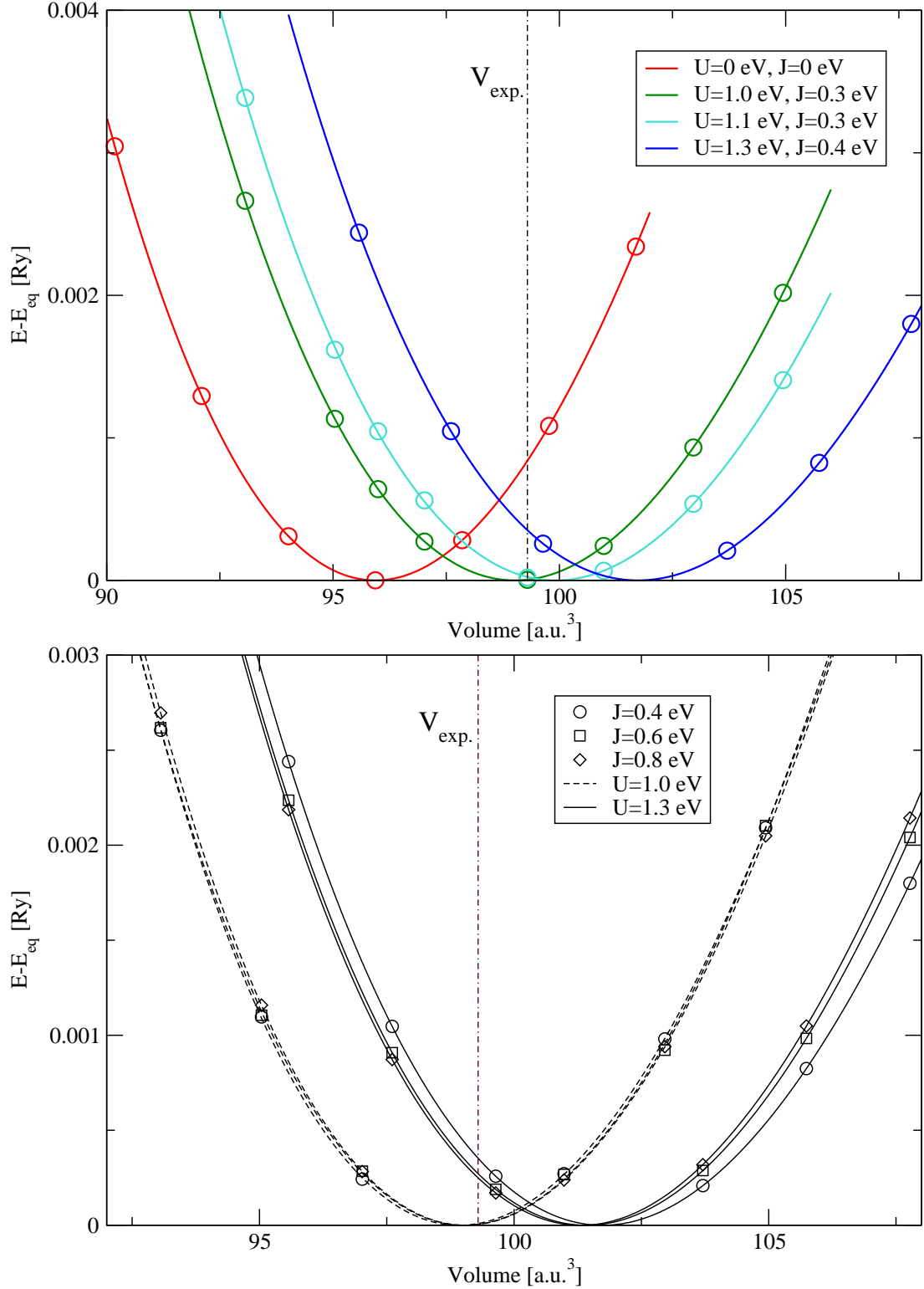

\includegraphics[scale=0.6,clip=true]{fig1.eps}\\
\includegraphics[scale=0.6,clip=true]{fig2.eps}
\caption{(Color online) Equation of state curves. Top: Effect of increasing $U$. LDA (red) compared to $U=1.0$ eV, $J=0.3$ eV (green); $U=1.1$ eV, $J=0.3$ eV (turquoise); $U=1.3$ eV, $J=0.4$ eV (blue). Bottom: Effect of altering $J$ while keeping $U$ fixed, for $U=1.0$ eV (dashed line) and $U=1.3$ eV (solid line).}
\label{eoscurves}
\end{figure}

\begin{table*}
\caption{Experimental lattice constants $a$ (and equivalent unit cell volume) of palladium 
from various sources, as function of temperature.}\label{tab2}
\begin{ruledtabular}
\begin{tabular}{ccccl}
$T$ [K] & $a$ [{\AA}] & $a$ [a.u.] & Volume [a.u.$^3$] & \\
\hline
853 & 3.9184 & 7.4047 & 101.50 & Ref.~\onlinecite{mi.br.71} \\
673 & 3.9088 & 7.3866 & 100.76 & Ref.~\onlinecite{mi.br.71} \\
297 & 3.9049 & 7.3792 & 100.45 & Ref.~\onlinecite{ra.ra.64} \\
296 & 3.8904 & 7.3518 & 99.34 & Ref.~\onlinecite{mi.br.71} \\
296 & 3.8902 & 7.3514 & 99.32 & Ref.~\onlinecite{mi.br.71} \\
120 & 3.8830 & 7.3378 & 98.77 & Ref.~\onlinecite{mi.br.71} \\
23 & 3.8907 & 7.3524 & 99.36 & Ref.~\onlinecite{ra.ra.64} \\
0\footnotetext[1]{Estimated from room temperature using linear thermal expansion coefficient, see Ref.~\onlinecite{st.sc.04}}\footnotemark[1] & 3.881 & 7.334 & 98.62 & Ref.~\onlinecite{st.sc.04} \\
0\footnotetext[2]{Corrected for zero-point anharmonic expansion, see Ref.~\onlinecite{st.sc.04}}\footnotemark[2] & 3.877 & 7.326 & 98.32 & Ref.~\onlinecite{st.sc.04} \\
\end{tabular}
\end{ruledtabular}
\end{table*}

\begin{table*}
\caption{Equilibrium volumes $V_0$ and bulk modulii $B_0$ extracted from equation of state fitting function (Birch-Murnaghan), for different sets of $U$ and $J$ parameters. The experimental volume $99.3$ a.u.$^3$ is taken from the room-temperature data of Ref.~\onlinecite{mi.br.71}, which differs from the $T=0$ K data by $<1\%$. The experimental bulk modulus is 189 GPa~\cite{young.91}.}\label{tab1}
\begin{ruledtabular}
\begin{tabular}{cccc}
$U$ [eV] & $J$ [eV] & $V_0$ [a.u.$^3$] & $B_0$ [GPa] \\
0 & 0 & 95.94 & 226.6 \\
\hline 
1.0 & 0.3 & 99.02 & 190.6 \\
    & 0.4 & 98.92 & 192.2 \\
    & 0.6 & 99.03 & 192.2 \\
    & 0.8 & 99.05 & 193.2 \\
\hline 
1.1 & 0.3 & 99.92 & 181.7 \\
\hline 
1.3 & 0.4 & 101.74 & 167.7 \\
    & 0.6 & 101.42 & 171.9 \\
    & 0.8 & 101.31 & 174.7 \\
\end{tabular}
\end{ruledtabular}
\end{table*}

We begin our study by showing that our LDA+DMFT method can accurately determine the equilibrium lattice constant and bulk modulus, two important ground state properties. The Coulomb and exchange parameters $U$ and $J$ that are to be used in the DMFT calculations are considered as adjustable parameters in this study, but can in principle be calculated within a first-principles framework \cite{sa.fr.11}. In this section we have decided to adjust the $U$ and $J$ values until the calculated equation of state (EOS) energy-volume curve reproduces the experimental lattice constant (see Table \ref{tab2} for a collection of experimental lattice constants from the literature).

In Figure \ref{eoscurves} (top) EOS curves for different values of $U$ and $J$ are presented. The experimental volume has been marked out. The equilibrium volume $V_0$ and bulk modulus $B_0$ for each of the curves can be seen in Table \ref{tab1}. It is seen (Figure \ref{eoscurves}, top) that $U=J=0$ eV (red curve), i.e. the LDA, underestimates the volume, which is commonly known. The generalized gradient approximation (GGA) to the exchange-correlation potential, as pointed out for Pd in Ref.~\onlinecite{al.ma.06}, overestimates the lattice constant, and leads to a ferromagnetic ground state and is therefore unsuitable. As the value of $U$ is increased, the computed lattice constant is increased towards the experimental value. For $U=1.0$ eV the calculated $V_0$ and $B_0$ for different exchange parameters $J$ are given in Table \ref{tab1}, and the values are closer to experiment than the LDA value. The effect on the EOS by varying the exchange parameter $J$ can be seen in Figure \ref{eoscurves} (bottom, dashed lines). The equilibrium volumes are tabulated in Table \ref{tab1}, and give a standard deviation of $0.05$ a.u.$^3$, which is of the same order as the scattering in the data for room temperature (See $T=296$ K in Table \ref{tab2}).
At $U=1.1$ eV and $J=0.3$ eV, $V_0$ is overestimated compared to the experimental value and $B_0$ is underestimated. Increasing
$U$ to 1.3 eV leads to an even larger $V_0$ and smaller $B_0$. Varying $J$ at this value of $U$ gives a standard deviation of $0.18$ a.u.$^3$, which is an order of magnitude larger then the standard deviation at $U=1.0$ eV. The effect of exchange $J$ on the volume is larger for $U=1.3$ eV than for $U=1.0$ eV, but it is still below the effect of the experimentally observed thermal expansion (see Table \ref{tab2}).
The increase of $J$ (for a fixed $U=1.3$ eV) decreases the equilibrium volume, which is opposite to the trend given by increasing $U$. However, this is a small effect and not relevant to this study.

Based on the results presented in this Section, $U=1.0$ eV and $J=0.3$ eV can be taken as reasonable choices in order to be able to reproduce the lattice constant and bulk modulus within our LDA+DMFT method.

\subsection{Ferromagnetic instability}
\begin{figure}[h!]
\includegraphics[scale=0.6]{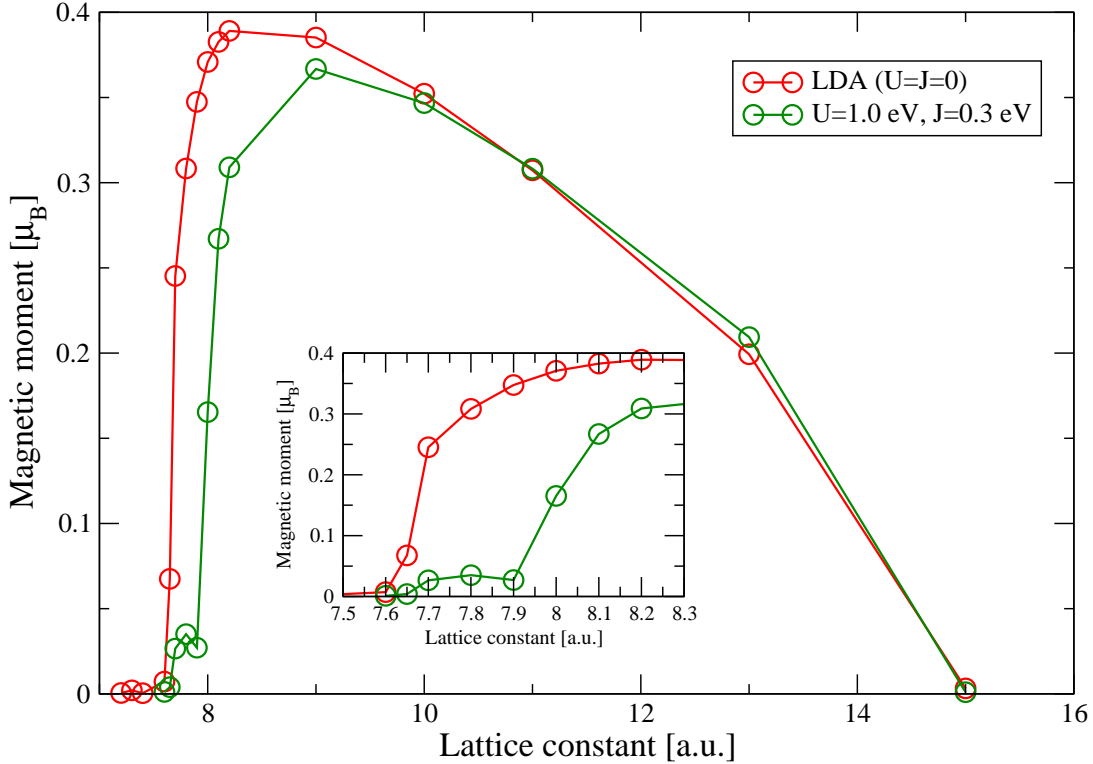}
\caption{(Color online) Magnetic moment calculated as a function of volume, within the LDA (red circles) and within LDA+DMFT (green circles) for $U=1.0$ eV and $J=1.3$ eV. Relativistic effects were treated using the scalar relativistic approximation.}
\label{mmlda}
\end{figure}
It is known that palladium is on the verge of ferromagnetism, having a large density of states at the Fermi level $D(E_F)$ leading to a large static susceptibility. An early theory that tried to explain the magnetic transition in itinerant electron systems was the Stoner model. According to this model, a magnetic state is favored over a non-magnetic state when the criterion $D(E_F)I \geqslant 1$ is fulfilled, where $I$ is the Stoner parameter~\cite{mohn_book}. 
This criterion points to the possibility of inducing magnetic order by increasing $D(E_F)$. In some cases, this can be achieved by reducing the effective dimensionality of the system. To create magnetic order attempts have been made to
lower the dimensionality of Pd systems, e.g by creating nanoparticles/wires~\cite{sa.cr.03,de.to.04,vi.jo.00} or thin films~\cite{sa.sa.14}. There also exists density functional theory studies that indicate that bulk palladium turns ferromagnetic as the volume is expanded \cite{fr.no.87,mo.ma.89,ch.br.89,ho.le.08}. 

In Figure \ref{mmlda} the magnetic moment in units of $\mu_B$ is plotted as a function of lattice constant. For the LDA within the scalar-relativistic approximation (red curve) a magnetic onset is brought about at a lattice constant of 7.65 a.u. This is $\sim 4\%$ larger than the experimental lattice constant, which is in accordance with previous studies, where the magnetic onset varies between a $1\%$-$6\%$ increase of the lattice constant. Hong and Lee \cite{ho.le.08} points out that this variance could be due to the sensitivity of $D(E_F)$ on the \textbf{k}-point mesh, and shows that $D(E_F)$ is difficult to fully converge even at dense mesh sizes. Note that the curve reaches a maximum ($\sim 0.4$ $\mu_B$) and then decreases toward zero magnetic moment at large lattice constants. A full charge transfer to the $d$-states has then been accomplished, leading to fully occupied $d$-states with no net magnetic moment~\cite{mo.ma.89}.

We next calculated the magnetic moment as a function of increasing lattice constant within the LDA+DMFT scheme, using the scalar-relativistic approximation, and setting $U=1.0$ eV and $J=0.3$ eV (Figure \ref{mmlda}, green curve). The magnetic transition is pushed further upwards in volume, compared to the scalar-relativistic LDA curve (red), giving a transition first into a ``low-moment" and then into a ``high-moment" state. We also note that the LDA+DMFT curve more or less coincide with the LDA curve at larger lattice constants. The system is then close to having a fully occupied $d$-band, where correlation should have negligible effect. 

DMFT is able to capture some dynamical spin fluctuation effects, and this could explain the suppression of the magnetic moment at those intermediate volumes where the LDA still gives noticeable moments.

\subsection{Density of states and Fermi surface}\label{resspec}
\subsubsection{Spectral functions and the formation of satellite structure}

\begin{figure}[h]
\includegraphics[scale=0.6]{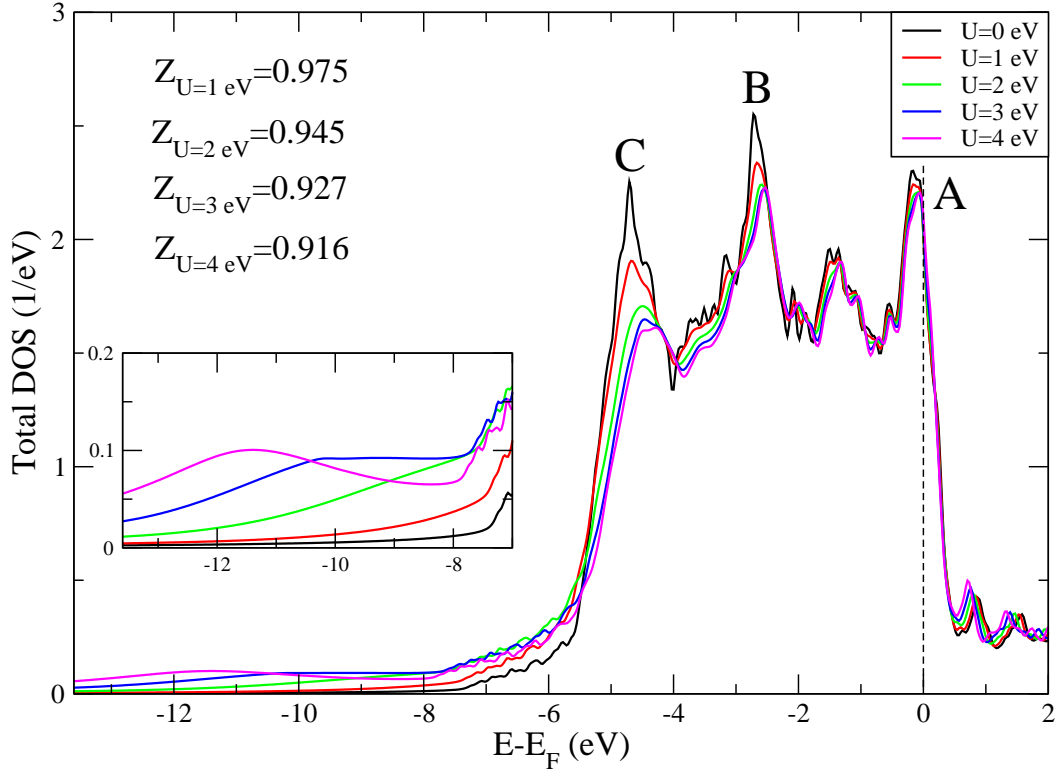}\\
\caption{(Color online) Total density of states as a function of the Coulomb interaction $U$. Note that
the peak closest to the Fermi level (marked by A) is pinned and that the lowest lying peak (C) decreases in intensity while a satellite
structure is formed for high binding energies (see inset). Corresponding quasiparticle weights $Z = (1-\partial Re[\Sigma(E)]/\partial E|_{E_F})^{-1}$ in the upper left corner.}
\label{fig2}
\end{figure}

Density of states (DOS) at the experimental lattice constant is presented in Figure~\ref{fig2}. Including electronic
correlations, for increased values of the local Coulomb parameter $U$
in the higher binding energy region a satellite structure develops.
We tuned $J$ for fixed $U$ and saw no significant change in DOS (not shown).
Hence, the satellite position is mostly insensitive to the value of the exchange parameter $J$.

The quasiparticle weights $Z = (1-\partial Re[\Sigma(E)]/\partial E|_{E_F})^{-1}$ for the different $U$ are reported in Figure \ref{fig2}, being in the range $Z=0.975-0.916$ for $U=1-4$ eV. These correspond to effective mass ratios $m^{*}/m_{LDA} = Z^{-1} = 1.03-1.09$, where $m_{LDA}$ is the LDA band mass. This should be compared with $m_{sp.heat}^{*}/m_{LDA} = 1.66$, where $m_{sp.heat}^{*}$ is estimated from electronic specific heat measurements and $m_{LDA}$ is taken from band structure calculations \cite{krogh.70,ha.sh.13}, which is considerably larger than what we get in this study. It should be noted that the electron-phonon coupling $\lambda_{e-ph}$ is not included in our self-energy, and previous theoretical studies have shown this quantity to be on the order $\lambda_{e-ph} \sim 0.35-0.41$ \cite{sa.sa.96,pi.al.78}. Recent angle-resolved PES (ARPES) by Hayashi et al.~\cite{ha.sh.13} estimated the electron-phonon coupling to be $\lambda_{e-ph} \sim 0.39$, and the electron-electron and electron-paramagnon coupling to be $\lambda_{e-e}+\lambda_{e-para} \sim 0.08$, leading to an effective mass $m_{ARPES}^{*}/m_{LDA} = 1+\lambda_{tot} \sim 1.5$. Using Hayashi et al.'s~\cite{ha.sh.13} value for $\lambda_{e-ph}$ and our calculated self-energy gives the effective mass $m^{*}/m_{LDA}=1.42-1.48$, for $U=1-4$ eV, which is in good agreement with Hayashi et al.~\cite{ha.sh.13}, but still underestimating the data from specific heat measurements.
It should be noted that our quasiparticle weights $Z$ are averaged over the BZ, while Ref.~\onlinecite{ha.sh.13} investigated specific paths in the BZ, while also being a surface sensitive study. This comparison however shows that our results are of similar magnitude.

\begin{figure}[h]
\includegraphics[scale=0.8]{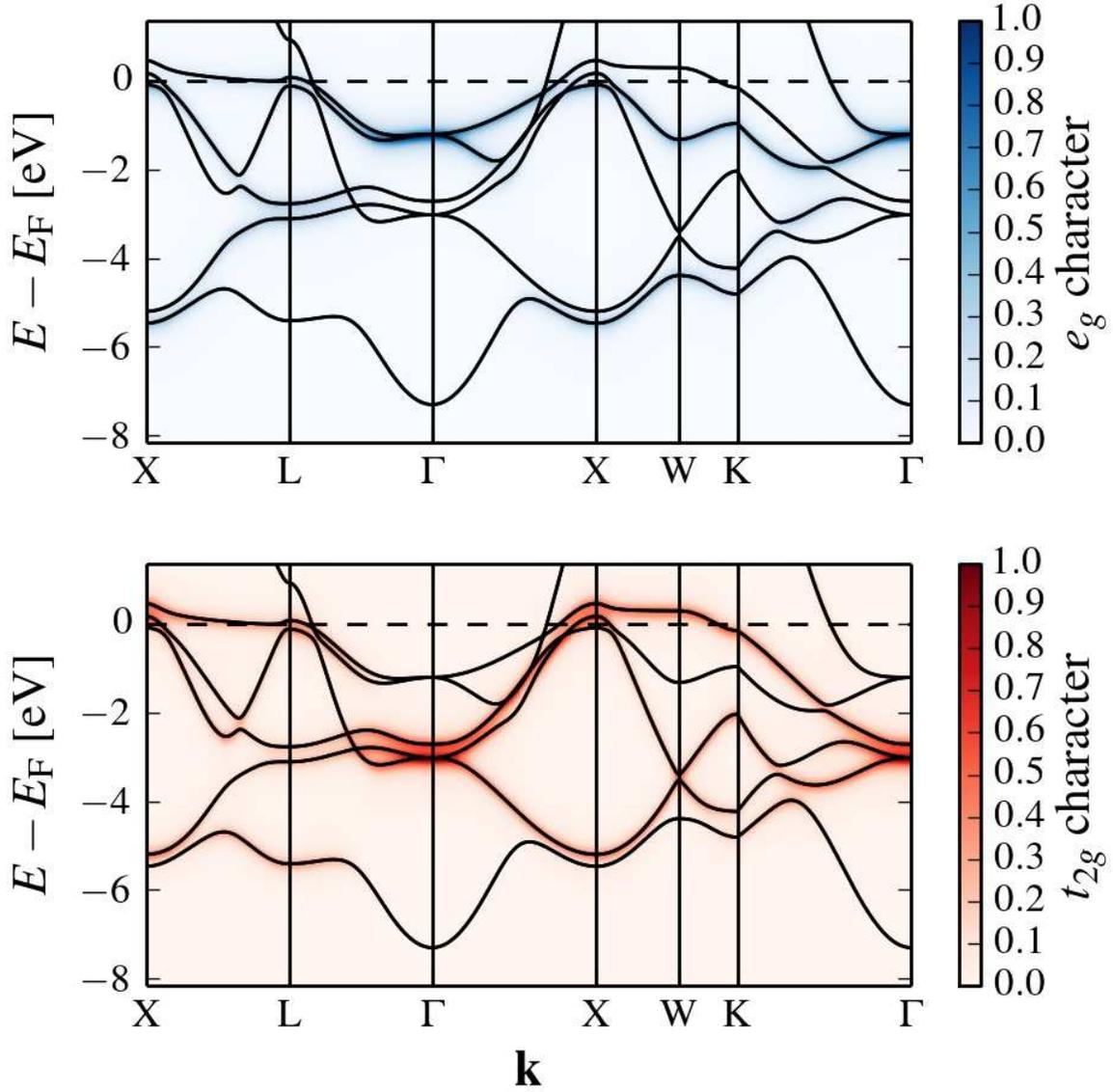}
\caption{(Color online) LDA orbital-resolved spectral functions along high symmetry lines in the BZ. Top: $e_g$-symmetry. Bottom: $t_{2g}$-symmetry.}
\label{fig7}
\end{figure}

Just below the Fermi level a dominant peak with a relatively 
large value of the density of states is situated at $\sim -0.15$ eV (marked by A) for all investigated $U$ values. A second major peak (B) is situated in the middle of the valence band at $\sim -2.7$ eV at $U=0$, and is shifted to $\sim -2.5$ eV as $U$ is increased. The third major peak (C) is at the bottom of the $d$-band at $\sim -4.7$ eV, and is shifted to $\sim -4.4$ eV as correlation is increased. The contributions of different bands to the peaks in the DOS can be inferred by studying the spectral function along high symmetry lines in the BZ, see Figure \ref{fig7}. 

Concerning the high-energy binding region in the photoemission spectra, there exist discrepancies on the order of $\sim 0.5$ eV between experiment and band structure calculations, as pointed out by Kang et al.~\cite{ka.hw.97}. The LDA seems to overestimate the bandwidth of Pd compared to the measured PES bandwidth, and some experimental states are located closer to the Fermi level than the theoretical states \cite{ka.hw.97,ll.qu.77,hi.ea.78,ya.hi.90}. It was proposed \cite{ka.hw.97} that surface and correlation effects could modify the LDA band structure, explaining the discrepancies. It is not altogether clear how to separate these two effects from each other since both bulk and surface states will contribute to the PES, especially for low photon energies. Kang et al. \cite{ka.hw.97} performed a combined PES and LDA level band structure calculation study for Pd, and their results indicated that surface effects could indeed explain the bandwidth narrowing. However, they also ruled out many-body correlation effects since they found no trace of a satellite in the PES. The missing satellite might be due to the neglect of the $4p-4d$ photoabsorption
threshold in Ref. \cite{ka.hw.97}, since the energy range of interest (around $\sim 55$ eV
photon energy) does not seem to be investigated. The experimental photoemission studies
in Refs. \cite{ch.kr.81,ni.la.81} scan this range
and do indeed find a satellite. The $4p-4d$ photoabsorption process can be viewed as
follows: A photon with energy at the $4p$ core level will excite a core electron
to the Fermi level. As the $4p$ core hole is filled by a valence electron, the resulting
valence hole will interact with the photoabsorbed electron and contribute to the satellite 
intensity. Note that the $4p-4d$ photoabsorption will affect the satellite intensity, but
not its position \cite{hufner}. The satellite position will be determined by the valence hole spectral
function, which we access in our calculations. We can not capture the 
contribution from the core levels on the spectra, and hence the satellite intensity
we obtain should not be directly compared with experiment. From comparison with Figure \ref{fig2}
and the experimental satellite position 8.5 eV \cite{ch.kr.81}, the $U$-value needed to reproduce
the satellite position can be estimated to be between 2-3 eV. By including correlation we also get
a shift of the B and C peaks to lower binding energy, in better agreement with experiment. 
The B peak position has been measured at $-2.55$ eV (Ref.~\onlinecite{hi.ea.78}), $-2.4$ eV (Ref.~\onlinecite{ni.la.81})
and $\sim -2.5$ eV (estimated from Ref.~\onlinecite{ka.hw.97}), which indicates that the LDA positions this peak at too
high binding energy ($\sim -2.7$ eV in this study) and that including correlation will improve the peak position
in comparison with experiment.
We here mention that no attempt was made to model the surface states, instead only
bulk calculations were performed. Note that matrix element effects were also not taken into
account in this study.

As shown in Section
\ref{reseos} a $U$-value above 1.0 eV would overestimate the equilibrium lattice constant. Hence,
to capture the experimental spectra, a different $U$ is needed than the
one that captures the experimental volume. That the $U$-value needed to reproduce
spectral features might differ from the value that reproduce the ground-state
properties has been observed also for Ni \cite{ma.mi.09,ka.li.02}, which makes
it plausible that similar behavior is observed for Pd.

It is interesting to discuss the satellite formation in Pd, in
comparison with Ni.
The effect of electron correlation on one-electron removal
energies from a partially filled band is described in terms of
interactions between three-body configurations, one hole
plus one electron-hole pair giving rise to hole-hole and hole-electron 
scattering~\cite{liebsch.79,liebsch.81}. The effectiveness of these scattering 
processes depends not only on the  strength of the screened
on-site electron-electron interaction, but also on the orbital
occupations involved in the scattering process. In particular 
on the number of empty $d$-states necessary for the creation of 
three-particle configurations since no electron-hole pair can 
be added to a completely filled band: in the case of nickel where 
only the minority-spin band has a sizable number of empty states
available the creation of a majority-spin hole will be followed 
by scattering processes involving only opposite spin electron-hole 
pairs. The interaction strength for this channel is of intensity 
proportional to $U$,  while the creation of a minority-spin hole 
will involve a scattering of strength proportional to $U-J$ with 
parallel spin electron-hole pairs only. In Pd both spin channels 
are symmetric at the equilibrium lattice parameter and fewer
 empty $d$-states are present in comparison with 
Ni. Despite the reduced scattering events in generating 
electron-hole pairs the existence of the $T$-matrix is enough 
in generating a satellite structure, although the scattering event is 
not very effective, since the satellite is hardly discernible for valence state 
spectroscopy. 

\begin{figure}[h]
\includegraphics[scale=0.5]{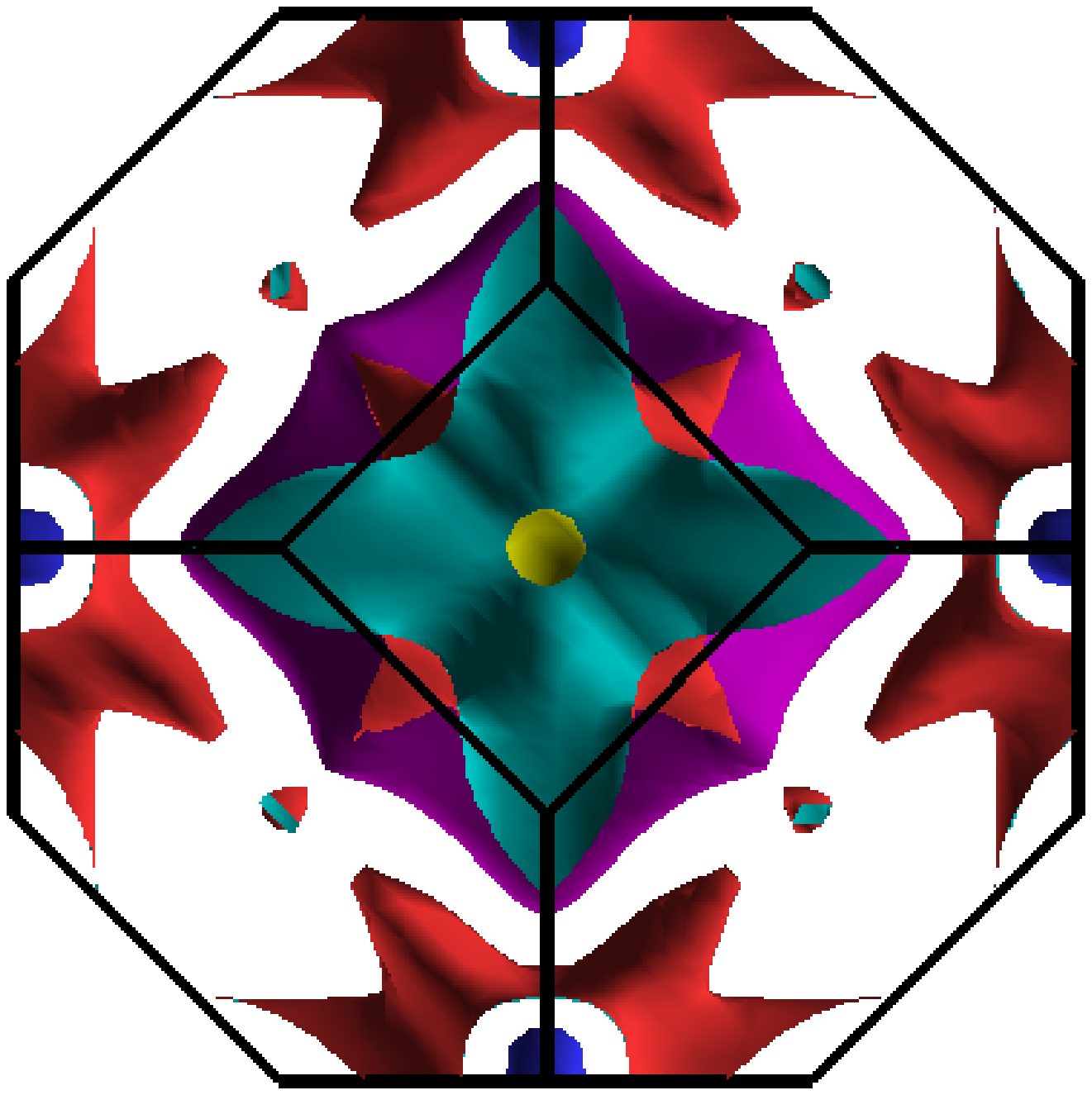}\includegraphics[scale=0.2]{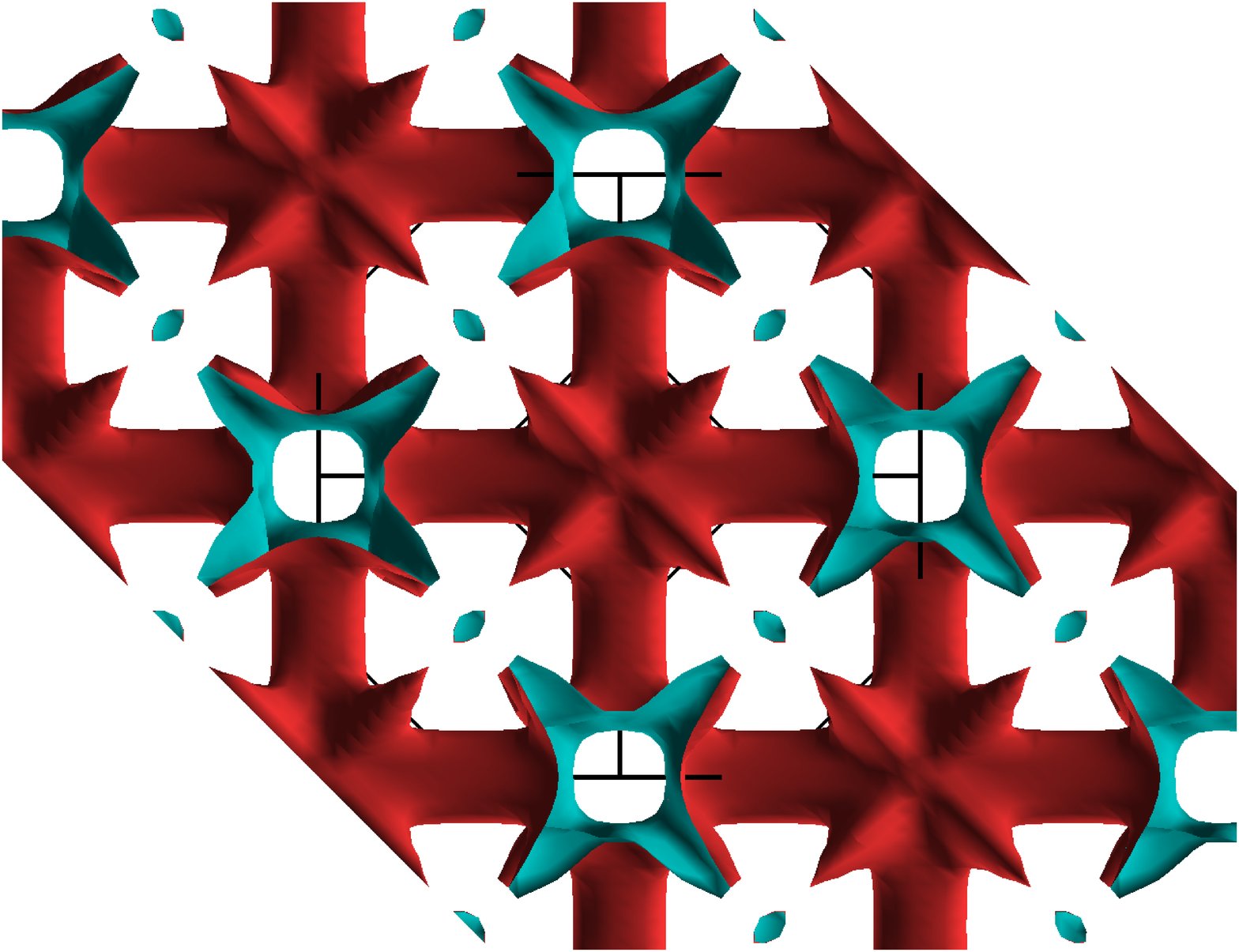}\\
\includegraphics[scale=0.4]{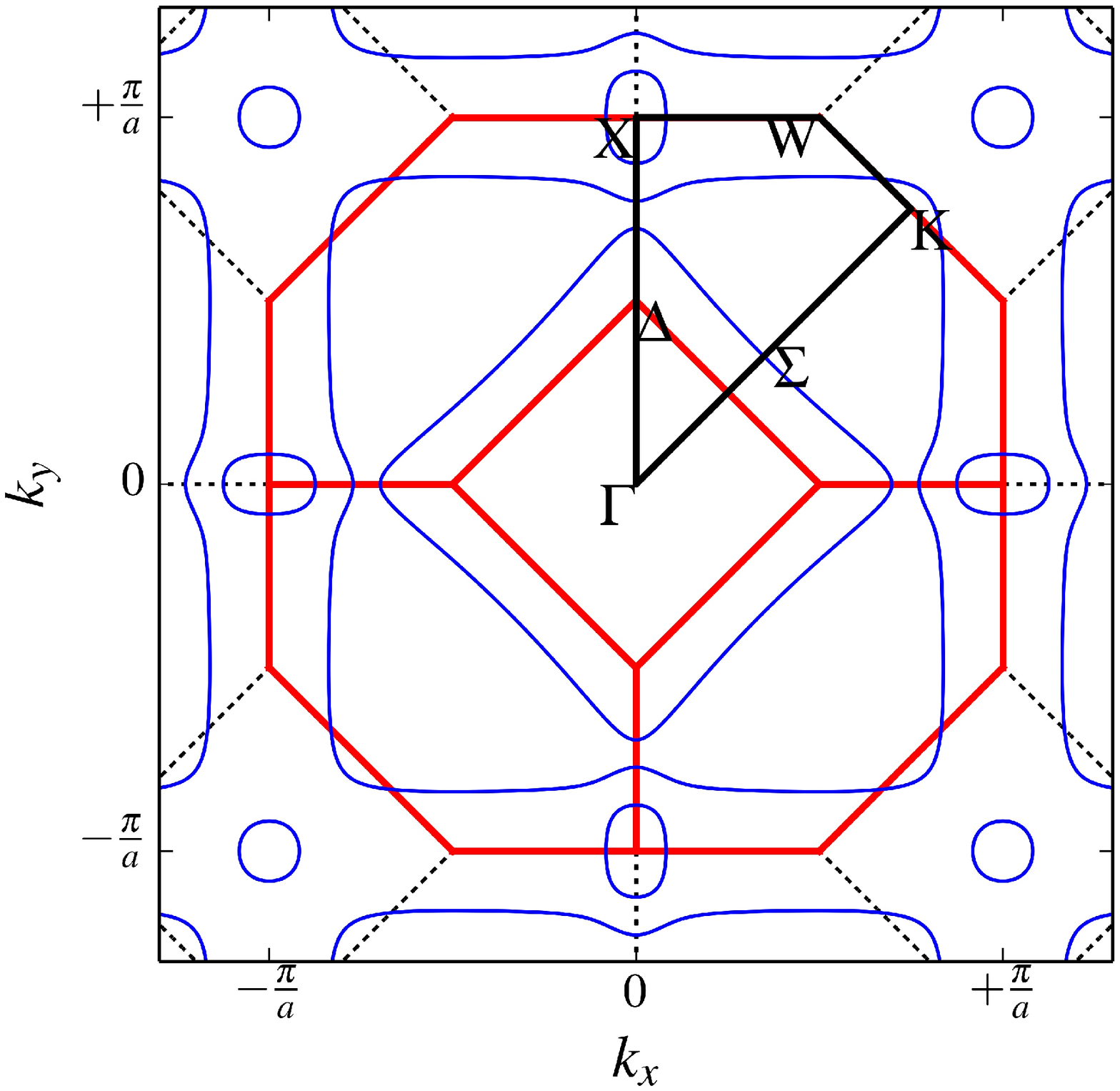}\includegraphics[scale=0.4]{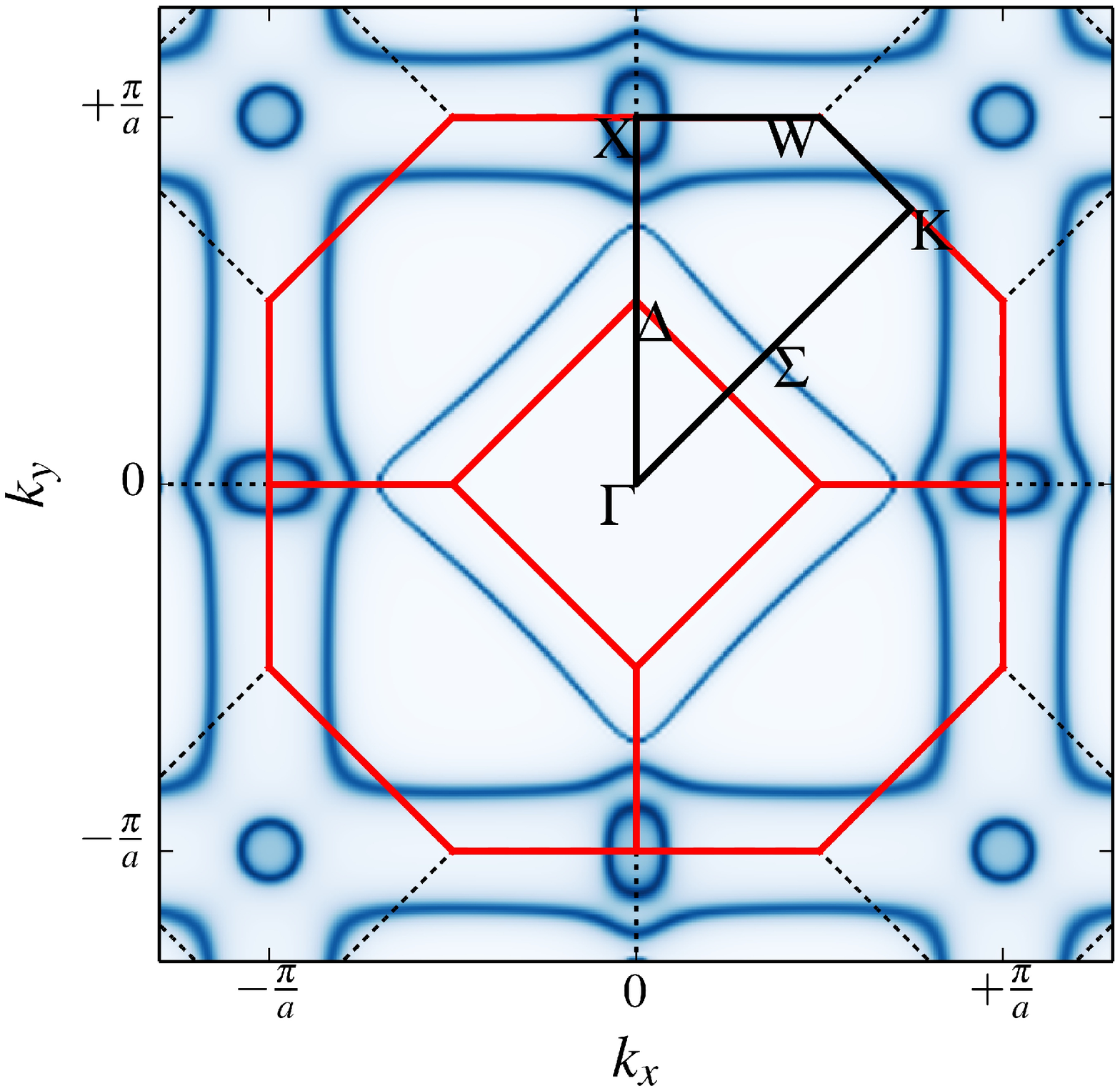}
\caption{(Color online) Fermi surfaces. Top, left: 3-dimensional Fermi surface in the first BZ, projected on the $k_x$-$k_y$ plane. Note the $X$ hole pockets centered at the square faces (hole side blue/electron side yellow), the $L$ hole pockets centered at the hexagonal faces (hole side red/electron side turquoise) and the tube hole structures intersecting at the $X$-points (hole side red/electron side turquoise). Also note that the $L$ pockets only exist if spin-orbit terms are included. A large electron surface sheet is centered around the $\Gamma$-point (purple). Top, right: Hole tube structure as seen in the extended zone scheme. Bottom, left: Cut at $k_z=0$ within the LDA. Bottom, right: Cut at $k_z=0$ within the LDA+DMFT, $U=1.0$ eV and $J=0.3$ eV. The 3-dimensional Fermi surface was created with the XCrysden software \cite{koka.03}.}
\label{fig3}
\end{figure}

\subsubsection{Fermi surface}
Many calculations for the Fermi surface of Pd
exist in the literature \cite{mu.fr.70,krogh.70,dy.ca.81}. We present a cut of the LDA Fermi surface
in the $k_x-k_y$ plane (Figure~\ref{fig3}, bottom left) together with a projection
 of the 3-dimensional FS sheets (Figure~\ref{fig3}, top left). The Fermi surface geometry contains the closed 
electron surface around the $\Gamma$ point, and a set of hole ellipsoids 
at the $X$ points. Open hole surfaces consists of cylinders, extending in the 
$[100]$ and $[010]$ directions (i.e. along the $X-W-X$ paths) and intersecting in pairs 
at the symmetry points $X$, see top right of Figure \ref{fig3}. The open hole surfaces are particularly interesting 
as they are associated with the large effective masses and contribute 
substantially at the density of states near the Fermi level~\cite{dy.ca.81}. The Kohn anomaly~\cite{kohn.59} in the slope of
the $[\xi \xi 0]$ transverse acoustic branch of the Pd phonon dispersion is attributed to 
Fermi surface nesting between these open hole cylinders (see Ref.~\onlinecite{stew.08}
and references therein). Previous calculations
also predicted the existence of small $L$-pockets, 
which were seen if spin-orbit coupling was
taken into account \cite{mu.fr.70,krogh.70}. 
These $L$-pockets were later confirmed by magnetoacoustic measurements~\cite{br.ka.72}.

The orbital character of the FS sheets can be determined by investigation of the orbital-resolved spectral
function, see Figure \ref{fig7}. The tube structure (stemming mostly from
the flat band between the $W$ and the $X$ symmetry points) has mostly $t_{2g}$ character, which
was pointed out already by Kanamori \cite{kana.63}.
Switching on correlation through DMFT, the FS can be seen in Figure \ref{fig3} (bottom right). No large
difference between the Fermi surface within the LDA is seen, and no topological transition occurs.
The Fermi surface nesting vector believed to be responsible for the Kohn anomaly seen in the
phonon dispersion of Pd is estimated to be $\mathbf{q}=\frac{2\pi}{a}[0.30,0.30,0]$, in close
agreement with previous studies~\cite{stew.08}.

We end this section by pointing out that the negligible change in diameter of the tube structure as
correlation is increased is reassuring. This is so since the Kohn anomaly is well captured already at
the level of the LDA \cite{stew.08}, and a change in radius would change also the FS nesting, destroying the
agreement with experiment. 

\subsection{Local and non-local correlation effects}\label{resloc}

\begin{figure}[h]
\includegraphics[scale=0.8]{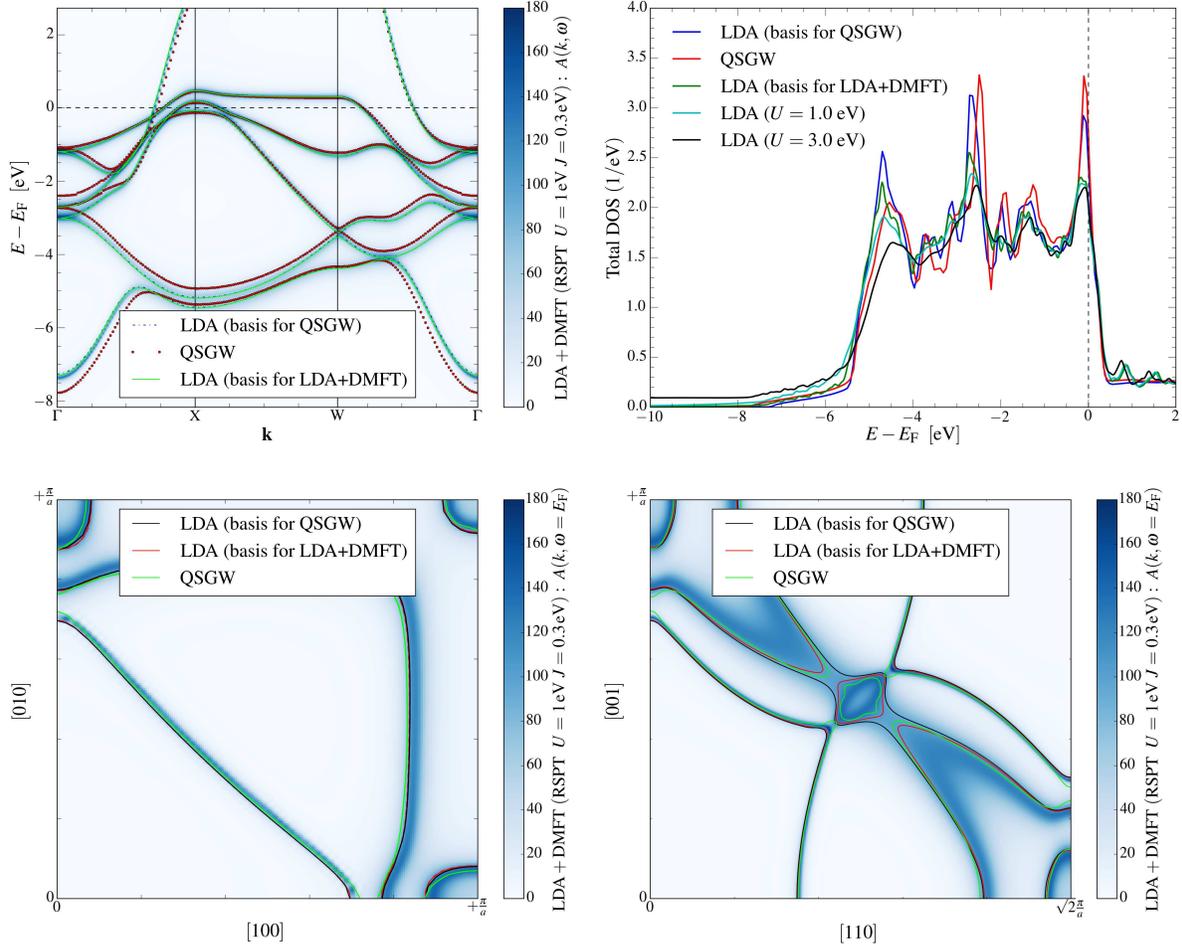}
\caption{(Color online) Blue color map corresponds to LDA+DMFT, $U=1.0$ eV and $J=0.3$ eV. Top left: Band structure along high symmetry directions in the BZ. Top right: QSGW and LDA+DMFT DOS. Bottom left: Fermi surface cut in $k_x-k_y$ plane. Bottom right: Fermi surface cut including $L$ point.}
\label{fig4}
\end{figure}

In order to investigate the effect of non-local electron correlations on the electronic structure of Pd, calculations were also performed using the QSGW method. Band structure; spectral functions and
Fermi surfaces were calculated using the experimental volume.

In Figure \ref{fig4} (top left) the band structure is plotted along high symmetry lines within the Brillouin zone. The
bands within the LDA from RSPt (green solid lines) and from QSGW (blue dashed lines) coincide well. Turning
on correlation effects, the bands are modified compared to the LDA result. The QSGW (red dots) and the LDA+DMFT (blue
energy scale) are seen to be nearly coinciding around the Fermi level, and differences are seen at higher energies. Around the $\Gamma$-point for energies between $-6$ eV and the Fermi level, the QSGW bands are shifted towards the Fermi level to a larger extent than the LDA+DMFT bands. For energies more negative than $-6$ eV, the lowest band is shifted downwards in energy to a larger extent than the LDA+DMFT bands. The trends (shift upwards/downwards in energy) is however the same for both methods, indicating that the $U$-value used in LDA+DMFT ($U=1.0$ eV) is too small to reproduce the correct quasiparticle eigenvalue position. This was also seen when comparing the LDA+DMFT spectral function with experiment in Section \ref{resspec}.

In Figure \ref{fig4} (top right) the DOS calculated within the QSGW method is plotted (red line). The DOS corresponding to the initial starting LDA solution is plotted in blue. The effect of correlation is most easily seen by inspecting the three main peaks in the DOS. In Figure \ref{fig4} (top right) we also show the LDA+DMFT \textbf{k}-integrated spectral function. The spectral functions within LDA+DMFT are calculated along a horizontal complex contour at a distance $\delta$ from the real axis, giving a broadening to the DOS. We performed LDA density of states calculations within RSPt along the real axis as well, and found excellent agreement with the LDA from QSGW (not shown). As correlation is turned on, similar trends in the three main peaks can be seen for the QSGW metod as was seen within the LDA+DMFT method. One main difference is that LDA+DMFT can produce the high energy satellite, while QSGW can not. This can attributed to the $T$-matrix ladder diagrams which are present in the LDA+DMFT self-energy, but not in the QSGW self-energy. There exist extensions of the $GW$ formalism that allow for $T$-matrix diagrams, see Refs. \cite{zh.ch.05,ro.be.12}.

In Figure \ref{fig4} (bottom left) the calculated Fermi surface in a cut of the $k_x-k_y$-plane from both LDA+DMFT and QSGW can be seen. Both of the two methods changes the FS slightly. The topology of the sheets is unchanged, but the \textbf{k}-space volume enclosed by the sheets shows some effect of correlation. The largest changes can be seen in the tube structure running along the $X-W-X$ symmetry directions. In the case of LDA+DMFT (blue intensity scale) the tube radius is slightly reduced, while for QSGW (green line) the radius is slightly increased. A different cut in the BZ, including
the $L$-pocket, is shown in Figure \ref{fig4} (bottom right). QSGW and LDA+DMFT display similar trends in the change of the FS, mainly  the beginning of a ``neck"-formation in the $\Gamma-L$ direction and a decreasing of the $L$-pocket diameter. Note that within the LDA solution used as a starting point for the QSGW, the $L$-pocket and the ``tongue" feature are connected along
the $X-L-X$ direction. We found that this was attributed to the use of the tetrahedron \textbf{k}-point integration method, which pushes the hole sheet slightly upwards in energy, creating the connection.

To conclude this section, we note that non-local effects captured by the QSGW method on the spectral functions come close to our LDA+DMFT data.

\section{Conclusion}\label{conc}

Electron correlation is commonly assumed to affect the electronic structure
of the $3d$ elements to a larger degree than in the $4d$ elements due, in part,
to the difference in $d$-state bandwidth. By electronic structure calculations within
 a LDA+DMFT context, we could show that even though the LDA can provide a reasonable description
 of the electronic structure of Pd, correlation effects give important contributions
 to ground-state and spectral properties. We could improve
the equilibrium lattice constant and bulk modulus from that of the LDA, and on expansion of the lattice 
constant Pd was shown to be ferromagnetic with a magnetic moment suppressed by spin fluctuations. 
The spectral function calculated with LDA+DMFT supported a formation of a
satellite in the high-energy binding region, while at the same time improving the band positions in
comparison with experiment. The spectral function and the Fermi surface showed
no major difference between the LDA+DMFT and QSGW method, and in particular the
nesting vector in the $[\xi \xi 0]$-direction was only slightly changed from its LDA value.

Within
the presently investigated LDA+DMFT method spin fluctuation effects were shown to
influence the magnetic transition volume, pushing it to higher values than within
the LDA. These results could point to that spin fluctuations could be important also for
the case of low dimensional systems like surfaces, nanoparticles and epitaxial thin films
of Pd.
 
This study confirms the band narrowing and favors the satellite formation seen in experiment for Pd. 
 Previously the difference between
the PES and band structure calculations has been attributed to surface effects \cite{ka.hw.97},
but our results indicate that also correlation should be taken into account, as was pointed
out earlier based on empirical arguments \cite{ni.la.81,ma.jo.80}. The LDA+DMFT method should be able
to probe the effect of correlation on the PES on an \emph{ab initio} level, and
further studies in conjunction with bulk and surface sensitive PES should
hopefully make it possible to disentangle surface and correlation effects from each other.

By performing $GW$ calculations with a \textbf{k}-dependent self-energy,
we could investigate the effect of non-local correlations on the spectral
properties of Pd.
 A closer inspection into momentum dependence of other properties could be interesting. 
Especially interesting would be to look into momentum dependent susceptibilities, which would be needed
to correctly address paramagons, which were recently observed in Pd~\cite{do.ha.10}. 

\section*{Acknowledgements}
We gratefully acknowledge financial support from the Deutsche Forschungsgemeinschaft through the Research Unit FOR 1346. I. D. M. and W. S. acknowledge financial support from the Swedish Research Council (VR), the Swedish strategic research programme eSSENCE and the Knut and Alice Wallenberg foundation (KAW, grants 2013.0020 and 2012.0031). M. R. also acknowledges support by the Ministry of Education, Science, and Technological Development of the Republic of Serbia under Projects No. ON171017 and No. III45018. M. S. acknowledges support by the Rustaveli national science foundation through the grant no. FR/265/6-100/14. L. V. acknowledges financial support from the Swedish Research Council and the Hungarian Scientific Research Fund (research projects OTKA 84078 and 109570). A. \"O. is also thankful for the financial support from the foundation of Axel Hultgren and from the Swedish steel producer's association (Jernkontoret). We acknowledge computational resources provided by the Swedish National Infrastructure for Computing (SNIC) at the National Supercomputer Centre (NSC) in Link\"{o}ping.

\end{document}